\documentclass[aps,prl,superscriptaddress,twocolumn]{revtex4}
\usepackage{graphicx}
\usepackage{amsmath}
\usepackage{bm}
\usepackage{xcolor}
\begin{document}

\title{Double Andreev Reflections in Type-II Weyl Semimetal-Superconductor Junctions
}

\author{Zhe Hou}
\affiliation{International Center for Quantum Materials, School of Physics, Peking University, Beijing 100871, China}
\author{Qing-Feng Sun}
\email[]{sunqf@pku.edu.cn}
\affiliation{International Center for Quantum Materials, School of Physics, Peking University, Beijing 100871, China}
\affiliation{Collaborative Innovation Center of Quantum Matter, Beijing 100871, China}

\begin{abstract}
We study the Andreev reflections (ARs) at the interface of the type-II Weyl semimetal-superconductor junctions
and find double ARs when the superconductor is put in the Weyl semimetal band tilting direction,
which is similar to the double reflections of light in anisotropic crystals.
The directions of the double (retro and specular) ARs are symmetric about the normal
due to the hyperboloidal Fermi surface near the Weyl nodes, but with different AR amplitudes
depending on the direction and energy of the incident electron.
When the normal direction of the Weyl semimetal-superconductor interface is changed from parallel
to perpendicular with the tilt direction,
the double ARs gradually evolve from one retro-AR and one specular AR, passing through double retro-ARs,
one specular AR and one retro-AR, into one retro AR and one normal reflection,
resulting in an anisotropic conductance which can be observed in experiments.
\end{abstract}

%\pacs{73.40.-c, % Electronic transport in interface
%74.50.+r,   % Tunneling phenomena
%74.45.+c,    % Andreev reflection (superconductivity)
%74.78.Na    % Mesoscopic and nanoscale systems
%}

\maketitle

In condensed matter physics, massless Dirac fermions or chiral Weyl fermions exist
as low-energy excitations and exhibit many novel phenomena in quantum
transport \cite{Graphene, add1,DiracFermion1, DiracFermion2, WeylFermion,add2}, triggering much interest in exploring new quasi-particles in real quantum systems \cite{Kitaev, FuKane, ThreeComponent}.
Materials hosting Weyl fermions are called Weyl semimetals (WSMs),
where the crossing points of the conduction and valence bands are known
as the ``Weyl nodes" with conic-like spectrums around them.
The Weyl node acts like a topological charge in momentum space
with the charge sign corresponding to its charity.
WSMs with point-like Fermi surface are referred as the conventional or type-I WSMs.
But recent progresses show that the conic spectrum can be tilted or overtilted to
transform the WSM into the type-II one \cite{Nature},
where the Fermi surface near the Weyl nodes is hyperboloidal with a large density of states.
The dramatic spectrum tilt violates the Lorentz-invariance and
generates electron and hole pockets near the Weyl nodes,
resulting in some distinct phenomena, e.g. the magnetic breakdown,
field-selective anomalous optical conductivity and anomalous Hall
effect \cite{MagneticB1, MagneticB2, AnomalousC1, AnomalousC2, AnomalousC3, AnomalousH}.
Experimentally LaAlGe, WTe$_{2}$, and MoTe$_{2}$ have been confirmed as the type-II WSMs
by observing the topological Fermi arcs between electron and hole pockets
using angle-resolved photo-emission spectroscopy (ARPES) \cite{LaAlGe, MoTe1, MoTe2, MoTe3, WTe1, WTe2}.

In this Letter, we investigate the type-II WSM-superconductor junctions
and find a new phenomena, double Andreev reflections (ARs), at the interface
caused by the spectrum tilt.
In the conductor-superconductor interface, except for the normal electron reflection,
there also exists AR \cite{Andreev},
a process where the electron is reflected back as a hole in the conductor
and a Cooper-pair is injected into the superconductor.
At the small bias, the conductance of the conductor-superconductor junction
is mainly determined by the AR \cite{add3}.
ARs can be classified into retro-AR and specular AR according to the directions of the reflected holes.
Retro-AR occurs in the normal metal-superconductor interface
where the hole almost retraces the path of the incident electron [Fig.1(a)]
and the electron-hole conversion is intraband,
i.e. both electron and hole locate in the same conduction or valence band.
In the graphene-superconductor junctions, the electron and reflected hole can locate in different bands,
of which the interband conversion results in the specular AR [Fig.1(b)] \cite{Beenakker}.
In both cases only single AR (retro or specular AR) occurs at the interface,
and the normal reflection exists usually.
However, in type-II WSM-superconductor junctions, the spectrum tilt makes
the conduction and valence bands lie in the same Fermi level,
so both intraband and interband electron-hole conversions happen
if the superconductor is in the tilt direction ($+x$ direction),
leading to the double ARs (retro and specular ARs) for one incident electron [Fig.1(c)].
The band tilt also forbids the normal reflection, resulting in a conductance
plateau when the bias is within the superconductor gap.
When the interface orientation of the WSM-superconductor junction is
changed from $+x$ direction to $+y$ direction,
the double ARs gradually evolve into one retro-AR and one normal reflection
with a decreasing subgap conductance,
indicating the ARs or conductance through type-II WSM-superconductor junctions are anisotropic.
We impose the anisotropic conductance as a useful detection for the tilting direction of the Weyl cones.

Consider a two-band effective Hamiltonian near the Weyl node $\textbf{\emph{K}}_{0}$ which respects the
time-reversal symmetry but breaks inversion symmetry:
\begin{equation}
  H_{+}(\emph{\textbf{k}})={\hbar}v_{1}k_{x}\sigma_{0}+{\hbar}v_{2} {\textbf{\emph{k}} {\cdot}\bm{\sigma}},
\end{equation}
where the wave vector $\textbf{\emph{k}}$ is the displacement from the Weyl node $\textbf{\emph{K}}_{0}$,
$\sigma_{0}$ is the identity matrix, $\bm{{\sigma}}$ is the Pauli matrix vector,
and $v_{2}$ is the Fermi velocity with its sign determining the charity of the node.
The spectrum tilt is described by the parameter $v_{1}$.
Here we set $v_{1}, v_{2}$ to be positive, and $v_{1}>v_{2}$ to make the WSM being the type-II one.
Time reversal symmetry requires another Weyl node locating at $-\textbf{\emph{K}}_{0}$ with its effective Hamiltonian satisfying \cite{TimeReversal}:
\begin{equation}
  H_{-}(\emph{\textbf{k}})=-{\hbar}v_{1}k_{x}\sigma_{0}-{\hbar}v_{2} (k_{x}{\sigma}_{x} -k_{y}{\sigma}_{y}+k_{z}{\sigma}_{z}).
\end{equation}
By substituting $\textbf{\emph{k}}$ with ${\textbf{\emph{p}}}/{\hbar}=-i{\nabla}_{\textbf{\emph{r}}}$, we get the effective Hamiltonian for Weyl nodes at $\textbf{\emph{K}}_{0}$ and $-\textbf{\emph{K}}_{0}$ in the real space: $H_{+}(\textbf{\emph{p}}/\hbar)$ and $H_{-}(\textbf{\emph{p}}/\hbar)$.

\begin{figure}
\includegraphics[width=8.4cm,clip=]{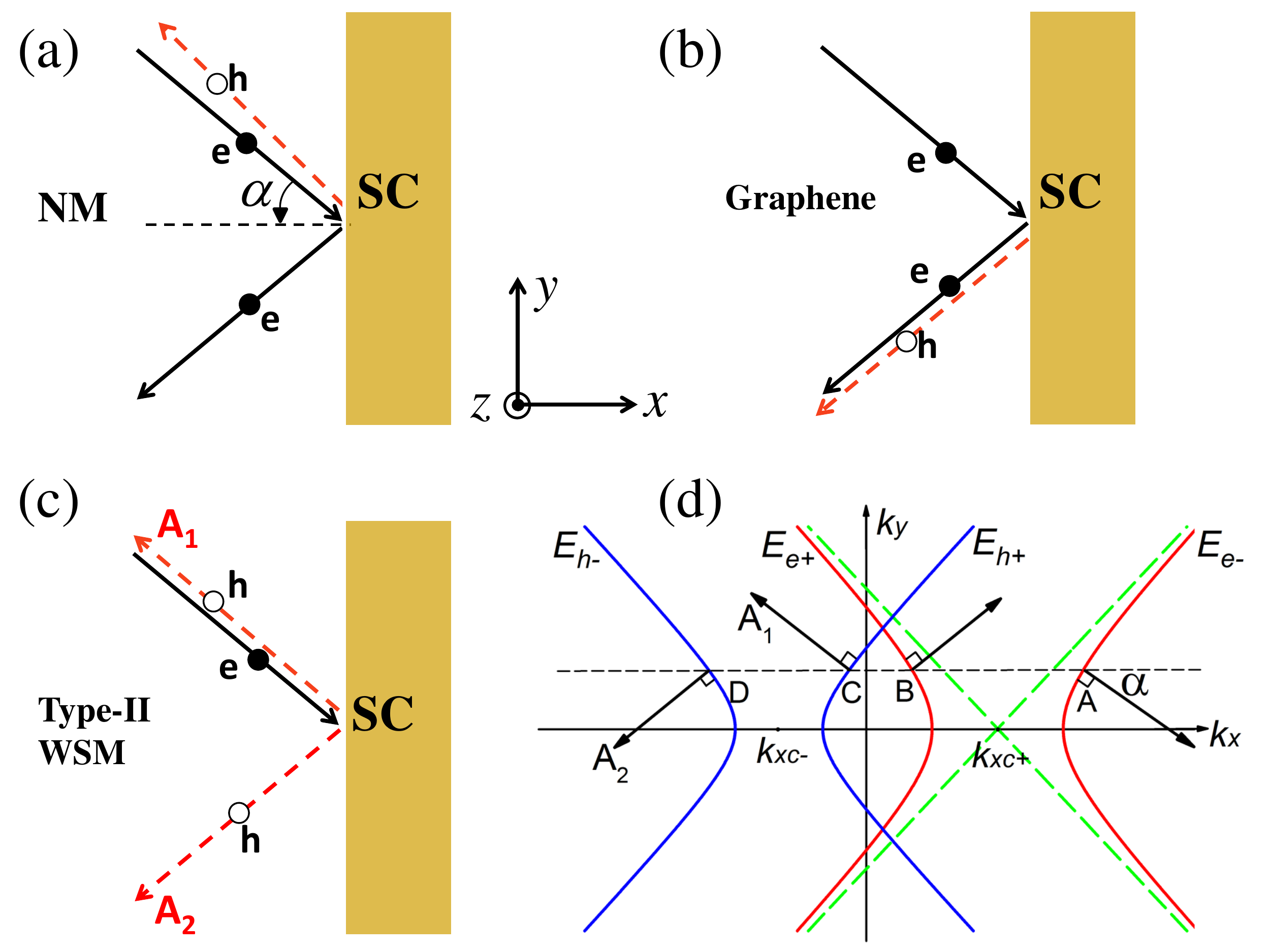}
\caption{(Color online) Schematic diagrams for: (a) retro-AR in normal metal-superconductor (SC) interface,
(b) specular AR in graphene-SC interface, and (c) double ARs in type-II WSM-SC interface.
The solid black arrows indicate the directions of the incident (reflected) electrons
and the dashed red arrows indicate the directions for the reflected holes.
(d) Hyperbolic equienergy lines ($E_{e\pm}, E_{h\pm}$) for electrons and holes in the $k_{x}$-$k_{y}$ plane.
The green dashed lines are the asymptotes for $E_{e\pm}$.
The intersections A, B, C, D denote
the incident modes for electrons and reflected modes for holes.
The black arrows at A and B denote the directions of the incident electron,
and the arrows at C and D are the directions for the double ARs. $\alpha$ is the incident angle.
}
\end{figure}

We consider the BCS pairing in the superconductor. Electron excitations near $\textbf{\emph{K}}_{0}$ are coupled by hole excitations near $-\textbf{\emph{K}}_{0}$, so the Bogoliubov-de Gennes (BdG) Hamiltonian reads \cite{BdG}:
\begin{eqnarray}
  H_{BdG}=\left(
\begin{array}{cc}
H_{+}(\emph{\textbf{p}}/\hbar)-\mu(\textbf{\emph{r}}) & \Delta(\textbf{\emph{r}}) \\
\Delta^{*}(\emph{\textbf{r}})  & -H_{-}^{*}(\emph{\textbf{p}}/\hbar)+\mu(\textbf{\emph{r}})
\end{array}
\right),
\end{eqnarray}
where $\mu(\textbf{\emph{r}})$, $\Delta(\textbf{\emph{r}})$ are the chemical potential and the pairing potential. One can verify that $H_{-}^{*}(\emph{\textbf{p}}/\hbar)=H_{+}(\emph{\textbf{p}}/\hbar)$ because of the time reversal symmetry.
The chemical potential $\mu(\textbf{\emph{r}})$ and the pairing potential $\Delta(\textbf{\emph{r}})$
are step functions:
\begin{eqnarray}
 \mu(\textbf{\emph{r}})=\left\{
\begin{array}{ll}
\mu     \hspace{1mm}       & {x      <   -y \tan\theta  }\\
U       \hspace{1mm}     & {x     >   -y \tan\theta },
\end{array} \right.
\Delta(\textbf{\emph{r}})=\left\{
\begin{array}{ll}
0  \hspace{1mm}      & {x      <    -y \tan\theta }\\
\Delta      \hspace{1mm}      & {x     >   -y \tan\theta }.
\end{array} \right.
\nonumber
\end{eqnarray}
In the WSM region ($x<-y \tan\theta$),
the chemical potential $\mu$ is tuned by a gate voltage and the superconductor gap is zero.
On the other side, it is the superconducting region ($x>-y \tan\theta$)
with a nonzero superconductor gap $\Delta$ and a large chemical potential $U$ ($U\gg \mu, \Delta$).
Here $\theta$ is the intersection angle of
the normal of the WSM-superconductor interface with the $x$-bias.

First, let us consider the case of $\theta=0$.
In this case, the WSM and superconductor are in $x<0$ and $x>0$ regions, respectively.
By solving the eigenvalues of $H_{BdG}$, we obtain the energy dispersions for electron and hole excitations in the type-II WSM:
\begin{eqnarray}
E_{e\pm}(\emph{\textbf{k}})&=&\hbar v_{1}k_{x}\pm\hbar v_{2}|\emph{\textbf{k}}|-\mu,\\
E_{h\pm}(\emph{\textbf{k}})&=&-\hbar v_{1}k_{x}\pm\hbar v_{2}|\emph{\textbf{k}}|+\mu,
\end{eqnarray}
and the excitation energy in the superconductor:
\begin{eqnarray}
E_{s}(k)=\sqrt{\Delta^{2}+(\hbar v_{1}k_{x}\pm\hbar v_{2}|\emph{\textbf{k}}|-U)^{2}}.
\end{eqnarray}
with $|\emph{\textbf{k}}|=\sqrt{k_{x}^{2}+k_{y}^{2}+k_{z}^{2}}$.
Denote $E_{e+}$ ($E_{h-}$) as the conduction band for electron (hole),
and $E_{e-}$ ($E_{h+}$) for the valence band.
The band tilt makes the slopes in the $+x$ direction of the bands for electrons (holes)
all positive (negative), so there are two incident modes for electrons
and two reflected modes for holes. Here we consider the electron incident from the valence band ($E_{e-}$).
The holes are then reflected into the conduction band ($E_{h-}$) and the valence band ($E_{h+}$).
The intraband electron-hole conversion results in the retro-AR ($A_{1}$)
and the interband conversion is for the specular AR ($A_{2}$) [see Fig.1(c-d)].
So at the WSM-superconductor interface, a beam of incident electron
is Andreev reflected into two beams of holes,
which is similar to the double reflections of light in anisotropic crystals \cite{DR1, DR2}.

In the momentum space, the equienergy surfaces (Fermi surfaces) for electrons and holes are circular hyperboloids satisfying the equation
$
\frac{(k_{x}-k_{xc\pm})^2}{a_{\pm}^{2}}-\frac{k_{y}^{2}+k_{z}^{2}}{b_{\pm}^{2}}=1,
$
where $k_{xc\pm}=\frac{(\mu\pm E)v_{1}}{\hbar (v_{1}^{2}-v_{2}^{2})}$, $a_{\pm}=\frac{|\mu\pm E|v_{2}}{\hbar (v_{1}^{2}-v_{2}^{2})}$,
$b_{\pm}=\frac{|\mu\pm E|}{\hbar \sqrt{(v_{1}^{2}-v_{2}^{2})}}$, and `+' for electrons, `-' for holes.
During the reflection process, the energy $E$, the wave-vectors $k_{y}$ and $k_{z}$ keep invariant.
Since the equienergy surface has rotational symmetry along the $k_{x}$-axis, for simplicity we set $k_{z}=0$ in the following analysis.
Then in the $k_{x}-k_{y}$ plane, the equienergy lines become hyperbolas as shown in Fig.1(d).
Given $E$ and $k_{y}$,
the intersections A, B, C, D denote the modes for the incident electron and the reflected holes with their semiclassical velocity $(v_{x},v_{y})\equiv\nabla_{\textbf{\emph{\textbf{k}}}}E(\textbf{\emph{\textbf{k}}})/\hbar$ [$E(\textbf{\emph{k}})$ is the energy dispersion for electrons or holes] being perpendicular
with the equienergy lines.
The black arrow at point A denotes the direction for the incident electron
with an incident angle $\alpha$ ($\alpha=-\arctan v_y/v_x$),
and the arrows at C and D are the directions for retro-AR ($A_{1}$) and specular AR ($A_{2}$).
$E_{h+}$ and $E_{h-}$ is symmetric about $k_{x}=k_{xc-}$,
so $A_{1}$ and $A_{2}$ have opposite reflection angles.
With the increasing of $|k_{y}|$, $|\alpha|$ increases but saturates at a critical angle $\alpha_{c}=\arctan(v_{2}/\sqrt{v_{1}^{2}-v_{2}^{2}})$.

The AR coefficients $A_{1}$ and $A_{2}$ can be calculated by solving the eigenstates of $H_{BdG}$ and matching the wavefunctions at the interface $x=0$ (see the Supplemental Material for detailed calculations \cite{Supplementary Material}):
\begin{eqnarray}
A_{1/2}(E)
&=&\left|\frac{(k_{\pm}'\mp k_{z}) (v_{1} k_{\pm}'\mp v_{2}k_{x\pm}')}
{(k_{-}-k_{z})(v_{1}k_{-}-v_{2}k_{x-})}\right|
\left|r_{\pm}
\right|^{2},
\end{eqnarray}
where $k_{x-}/k'_{x+}/k'_{x-}$ is the $k_x$ coordinate of the point A/C/D in Fig.1(d),
$k_{-}/k'_{+}/k'_{-} =\sqrt{(k_{x-}/k'_{x+}/k'_{x-})^2 +k_y^2+k_z^2}$ and
\begin{eqnarray}
r_{\pm}=\frac{(k_{x-}+ik_{y})(k_{\mp}'\pm k_{z}) \pm (k_{-}- k_{z})(k_{x\mp}'+ik_{y})}
{(k_{+}'-k_{z})(k_{x-}'+ik_{y})+(k_{x+}'+ik_{y})(k_{-}'+k_{z})}
e^{-i\beta}
\nonumber
\end{eqnarray}
with $\beta =\arccos E/\Delta$ if $E<\Delta$, $\beta=-i\,\textrm{arcosh} E/\Delta$ if $E>\Delta$.
Note the specular-AR (retro-AR) coefficient has the same amplitudes
for the electron incident from the valence band $E_{e-}$ and conduction band $E_{e+}$ \cite{Supplementary Material}.
For the normal incidence ($\alpha=0$), the results reduce into
\begin{eqnarray}
A_{1}(E)=\Theta(\mu-E)T^A(E),\hspace{1mm}
 A_{2}(E)=\Theta(E-\mu)T^A(E),
\end{eqnarray}
where $\Theta$ is the Heaviside step function and $T^A(E)=1$ if $E<\Delta$,
$T^A(E)=(E-\sqrt{E^{2}-\Delta^{2}})^{2}/\Delta^{2}$ if $E>\Delta$,
so only one AR happens at the normal incidence and $A_{1}$, $A_{2}$ exchange at the chemical potential $\mu$.
While for a general incidence case with $\alpha\not=0$, both $A_{1}$ and $A_{2}$ are nonzero.
Importantly, the total AR coefficient $A\equiv A_{1}+A_{2} =T^A(E)$,
which only depends on the energy $E$ and the gap $\Delta$,
with nothing to do with the incident angle $\alpha$ and other parameters.
Within the gap ($E<\Delta$), $A=1$, which means the electron-hole conversion at the interface
happens with unit probability. This is because no modes
for the normal electron reflection exist, and the normal tunneling from
the WSM to the superconductor is also prohibited when $E<\Delta$.

\begin{figure}
\centering
\includegraphics[width=8.5cm, clip=]{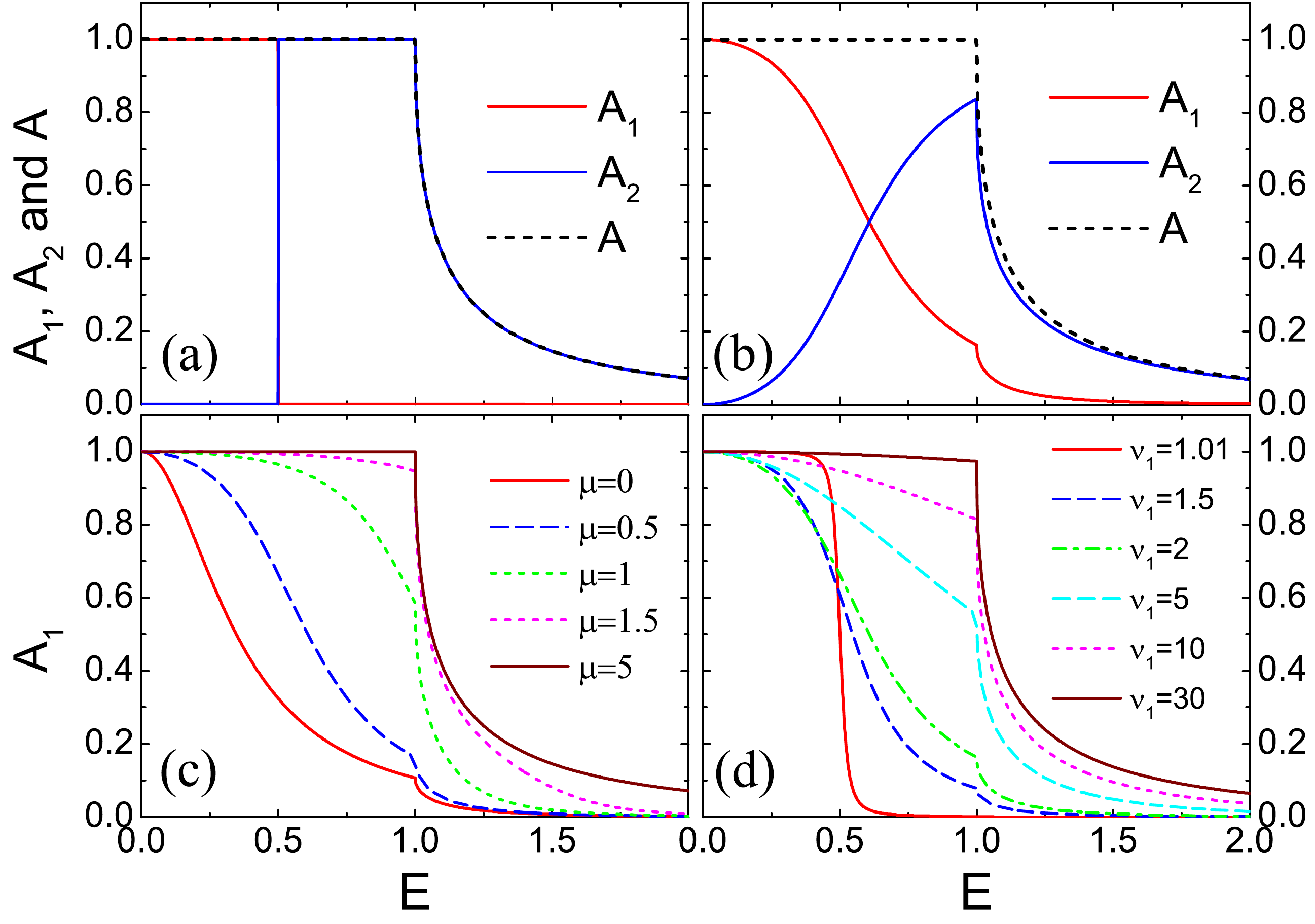}
\caption{(Color online) AR coefficients $A_{1}$, $A_{2}$ and $A$ as functions of the incident energy $E$. The parameters are: (a) $v_{1}=2$, $k_{y}=0$, $\mu=0.5$. (b) $v_{1}=2$, $k_{y}=0.2$, $\mu=0.5$. (c) $v_{1}=2$, $k_{y}=0.2$. (d) $k_{y}=0.2$, $\mu=0.5$.}
\end{figure}

In the following numerical calculations, we set $\hbar=1$, $v_{2}=1$, $\Delta=1$ and $k_{z}=0$.
Fig.2(a) shows AR coefficients $A_{1}$, $A_{2}$ and $A$ versus $E$ at the normal incidence ($\alpha=0$).
The results coincide with Eq.(8) and from the band structure point of view,
the modes symmetry for electron and hole plays an important role \cite{Supplementary Material}.
However, for the oblique incidence case ($\alpha\not=0$), both retro-AR and specular AR happen
[see Fig.2(b)].
When $E=0$, i.e. at the zero-energy incidence, we find $A_{1}=1$ and $A_{2}=0$.
As $E$ increases, $A_{2}$ goes up and $A_{1}$ falls down, and they both get a sharp
decline when $E>\Delta$.
The total AR coefficient $A$ shows the same variation behaviors in Fig.2(a) and (b)
because $A$ only depends on $E$ and $\Delta$.
In Fig.2(c) and (d) we change the chemical potential $\mu$ and the tilt parameter $v_{1}$
to see how $A_{1}$ varies. As $\mu$ increases, $A_{1}$ gets higher and approaches $A_{1}=T^A(E)$
in the large $\mu$ limit.
$A_{1}$ also shows the similar asymptotic behavior in the large tilting limit
[see $v_{1}=30$ in Fig.2(d)].
However, when the tilt is very small (see $v_{1}=1.01$), we find $A_{1}\simeq \Theta(\mu-E)$,
which is similar to the normal incident case.

The incident energy $E$ and angle $\alpha$ dependence of the AR coefficients $A_{1}$ and $A_{2}$
is investigated exhaustively in Fig.3.
The tilt $v_{1}$ is chosen to be $\sqrt{2}$,
then $\alpha$ has an up limit $\alpha_{c}=45^{\circ}$ from the analysis of Fig.1(d).
Below the chemical potential ($E<\mu$), $A_{1}$ first gets down and then goes up to $T^A$ with the increasing of $\alpha$ [Fig.3(a)].
While $A_{1}$ just increases monotonically with $\alpha$ when $E>\mu$.
For the ultimate angle incidence ($\alpha=\alpha_{c}$), $A_{1}=T^A$ and $A_{2}=0$, so only the retro-AR happens.
The specular AR $A_{2}$ shows the opposite behavior as $A_{1}$ due to the relation $A_{1}+A_{2}=T^A$.
Note that for appropriate $E$ and $\alpha$, the retro-AR and specular AR happen with nearly equal probabilities (see the green regions in Fig.3).
Another interesting thing we find is that for $\mu=0$,
$A_{1}$ and $A_{2}$ are independent of $E$ when $E<\Delta$ [Fig.3(c,d)],
which means once the incident angle $\alpha$ is given, the AR coefficients are determined.

Next we consider the nonzero orientation angle $\theta$ of the WSM-superconductor interface and
will show that the double ARs are anisotropic, strongly depending on $\theta$.
By performing the coordinate transformation:
\begin{eqnarray}
\left(
\begin{array}{cc}
\tilde{x} \\
\tilde{y}
\end{array}
\right)
=\left(
\begin{array}{cc}
\cos\theta & \sin\theta\\
-\sin\theta & \cos\theta
\end{array}
\right)
\left(
\begin{array}{cc}
x \\
y
\end{array}
\right),
\end{eqnarray}
the Hamiltonian of the WSM becomes:
\begin{eqnarray}
H_{+}&=&v_{1}\sigma_{0}(\tilde{p}_{x}\cos\theta-\tilde{p}_{y}\sin\theta)+v_{2}\sigma_{x}(\tilde{p}_{x}\cos\theta-\tilde{p}_{y}\sin\theta)\nonumber\\
&&+v_{2}\sigma_{y}(\tilde{p}_{x}\sin\theta+\tilde{p}_{y}\cos\theta),
\end{eqnarray}
where $\tilde{p}_{x}=-i\hbar\partial_{\tilde{x}}$, $\tilde{p}_{y}=-i\hbar\partial_{\tilde{y}}$.
The energy dispersion is:
\begin{eqnarray}
E_{e\pm}(\tilde{{\textbf{\emph{k}}}})=\hbar v_{1}(\tilde{k}_{x}\cos\theta-\tilde{k}_{y}\sin\theta)\pm \hbar v_{2}|\tilde{\textbf{\emph{k}}}|-\mu,
\end{eqnarray}
where $|\tilde{\textbf{\emph{k}}}|=\sqrt{{\tilde{k}_{x}}^{2}+{\tilde{k}_{y}}^{2}}$.
In the new coordinate system, the WSM and superconductor are respectively in $\tilde{x}<0$ and
$\tilde{x}>0$ regions, but the band tilts in both $\tilde{x}$ and $\tilde{y}$ directions.
Since $E$ and $\tilde{k}_{y}$ are conserved upon reflection at the interface $\tilde{x}=0$,
by performing the same calculation we obtain the directions and the amplitudes of
the double ARs \cite{Supplementary Material}.
Fig.4(a) is the schematic diagram of the evolutions of $A_{1}$ and $A_{2}$ when varying the
WSM-superconductor interface orientation angle $\theta$ at the fixed incident angle $\alpha=30^{\circ}$ and
incident energy $E=0.5$.

\begin{figure}
\includegraphics[width=8.8cm, clip=]{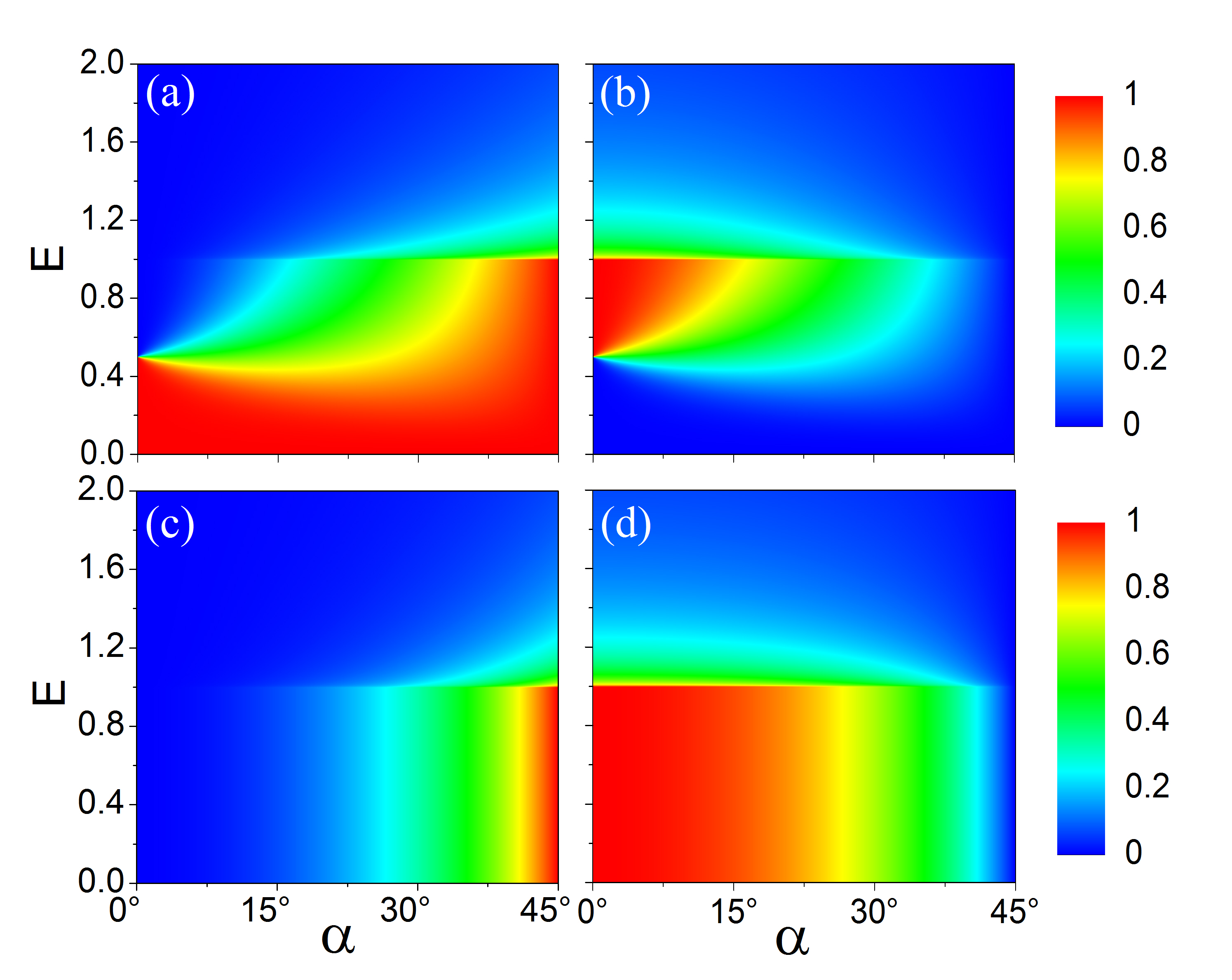}
\caption{(Color online) $E$ and $\alpha$ dependence of $A_{1}$ (a,c) and $A_{2}$ (b,d) with $v_{1}=\sqrt{2}$ and
$\mu=0.5$ (a,b) and $0$ (c,d).}
\end{figure}

\begin{figure}
\includegraphics[width=7.8cm,totalheight=2.65cm, clip=]{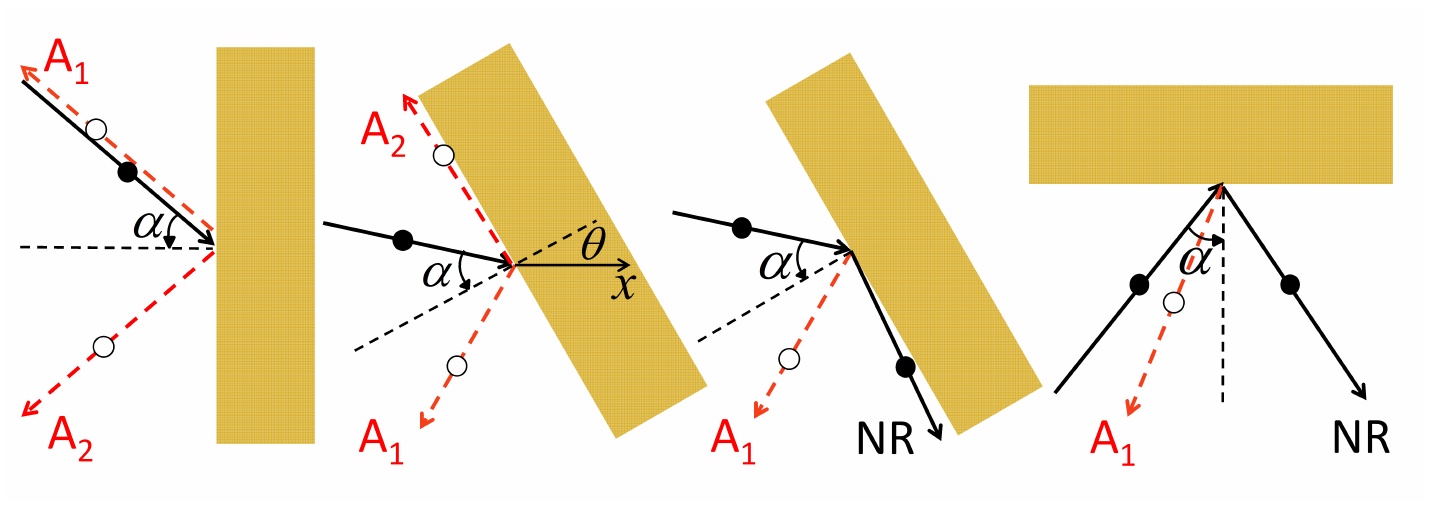}
\put(-235,65){(a)}

\includegraphics[width=8.0cm,totalheight=3.3cm, clip=]{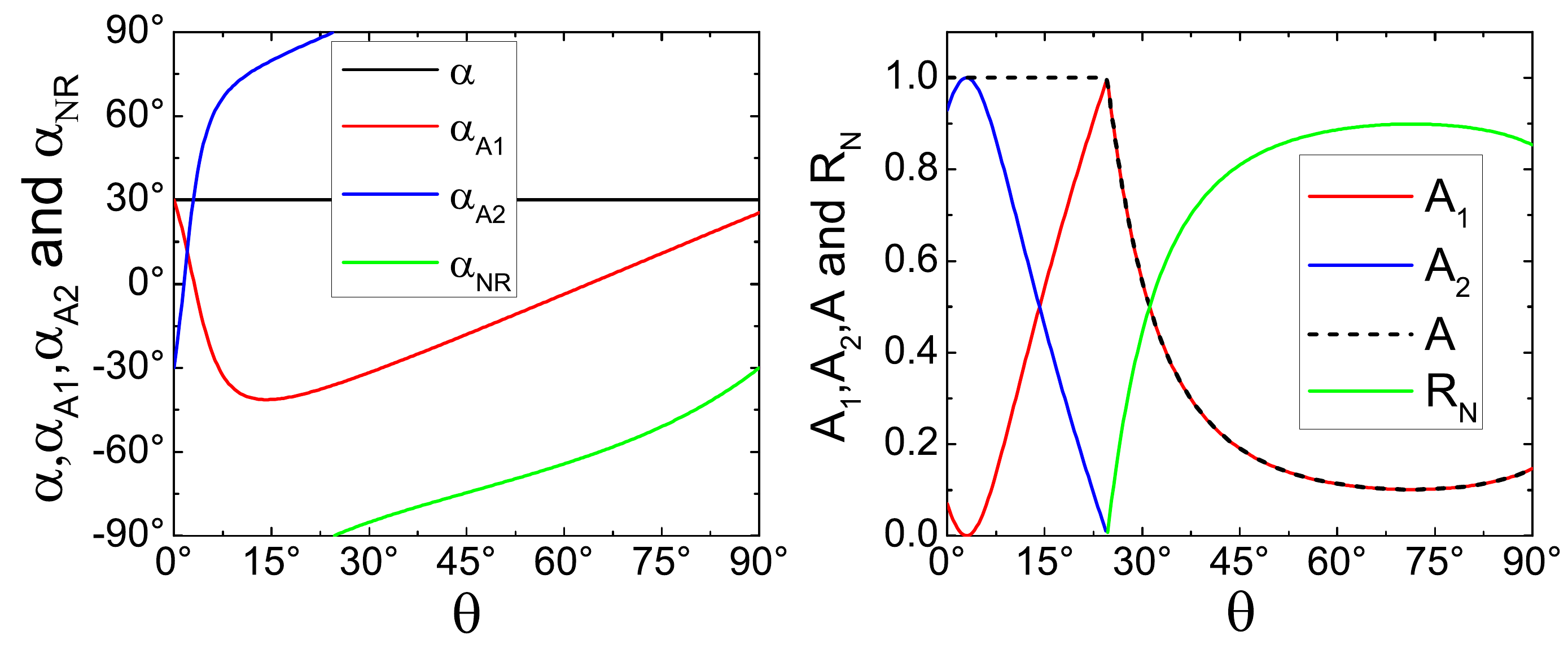}
\put(-238,90){(b)}
\put(-112,90){(c)}
\caption{(Color online) (a) Schematic diagrams of the AR $A_{1}$, $A_{2}$ and the NR
when changing the WSM-superconductor interface orientation angle $\theta$.
(b) The AR angles $\alpha_{A1}$, $\alpha_{A2}$ and NR angle $\alpha_{NR}$ as functions of $\theta$.
(c) The coefficients of $A_{1}$, $A_{2}$, $A$ and NR versus $\theta$.
Here the incident angle $\alpha$ is fixed to be $30^{\circ}$, the incident energy $E=0.5$, $\mu=0$ and $v_{1}=1.1$. }
\end{figure}

The corresponding reflection angles $\alpha_{A1}$ and $\alpha_{A2}$ are shown in Fig.4(b).
When $\theta=0$, the retro-AR $A_{1}$ has the same angle as the incident angle ($\alpha_{A1}=30^{\circ}$), and the specular AR $A_{2}$ is symmetric with $A_{1}$ with $\alpha_{A2}=-30^{\circ}$.
With the increasing of $\theta$, $\alpha_{A1}$ decreases and $\alpha_{A2}$ increases.
At $\theta \approx 2^{\circ}$, $\alpha_{A2}$ becomes positive,
indicating $A_{2}$ jumps into the other side of the normal and becomes the retro-AR.
Now double retro-ARs occur with both $\alpha_{A1}$ and $\alpha_{A2}$ being positive
while $2^{\circ} \lesssim \theta \lesssim 5^{\circ}$.
At $\theta \approx 5^{\circ}$, $\alpha_{A1}$ drops from positive to negative, leading
a retro-AR and a specular AR again. Note that now $A_2$ is retro-AR and $A_1$ is specular AR.
Further increasing $\theta$, $\alpha_{A2}$ goes up straightly to $90^{\circ}$,
then the AR $A_{2}$ disappears at the critical angle $\theta_{c}=\arccos{v_{2}/v_{1}}$
and the normal electron reflection arises with its reflection angle $\alpha_{NR}=-90^{\circ}$.
At the critical angle $\theta_{c}$, the double ARs evolve into one specular AR and
one normal reflection (NR) due to the band tilt in $\tilde{x}$ direction changing from the type-II ($v_{1}\cos\theta>v_{2}$) into the type-I ($v_{1}\cos\theta<v_{2}$).
Note that although the WSM still belongs to the type-II one,
only the tilt along the normal direction ($\tilde{x}$-axis) plays a decisive role in generating the double ARs.
With the further increasing of $\theta$ from $\theta=\theta_c$,
both $\alpha_{A1}$ and $\alpha_{NR}$ go up. Finally when $\theta=90^{\circ}$,
$\alpha_{NR}$ increases to $-30^{\circ}$ and
$A_{1}$ turns back to the retro-AR but with smaller AR angle $\alpha_{A1}$ than $\alpha$,
which is similar with the normal metal-superconductor junction.
Fig.4(c) shows the amplitudes for $A_{1}$, $A_{2}$ and the NR.
One can see that both $A_{1}$ and $A_{2}$ have large value with $A=A_{1}+A_{2}=1$ at $\theta<\theta_c$.
$A_{1}=1$ and $A_{2}=0$ at the critical angle $\theta_{c}$. Further increasing $\theta$,
$A_{2}$ disappears and the NR arises with its coefficient $R_{N}$ increasing with $\theta$
except for $\theta$ near $90^{\circ}$.
All the analysis above shows the spatial anisotropy of the double ARs.

The differential conductance of the type-II WSM-superconductor junction can be obtained from the BTK formula \cite{BTK}:
$G(eV)=\frac{2e^{2}S}{{\pi}^{2}h}\int d\tilde{k}_{y}\int dk_{z} [1+A(\tilde{k}_{y},k_{z},eV)-R_{N}(\tilde{k}_{y},k_{z},eV)]$,
with the cross-sectional area of the junction $S$. Here the spin and valley degeneracies \cite{Degeneracy}
are already considered.
While the interface angle $\theta<\theta_c$, $G(eV)$ can reduce into
$\frac{2e^{2}\tilde{q}^{2}S}{\pi h}[1+T^A(eV)]$, where $\tilde{q}$ is the cut-off value of the integral area
(i.e. $\tilde{k}_{y}^{2}+k_{z}^{2}\leq \tilde{q}^{2}$).
Within the superconductor gap $T^A(eV)=1$ and we find $G(eV)=\frac{4e^{2}\tilde{q}^{2}S}{\pi h}$,
which shows a plateau with respect to the bias $V$ and may be experimentally observed.
On the other hand, when $\theta>\theta_{c}$,
the conductance decreases and the plateau is broken,
because the NR arises and the total AR coefficient $A<1$ [Fig.4(c)] and depends on $eV$,
indicating the conductance through the type-II WSM-superconductor junction is strongly anisotropic.
This can be used to detect the tilt direction of the Weyl cones in real experiments.

In conclusion, we find the double ARs in the type-II WSM-superconductor junction.
The band tilt of the type-II WSM makes both interband and intraband electron-hole conversions happen
at the interface, resulting in the specular AR and retro-AR for one incident electron.
The differential conductance through the junction is also considered,
which exhibits strongly spatial anisotropy when varying the WSM-superconductor interface orientation.

Z. Hou thanks Peng Lv and Yan-Feng Zhou for helpful discussions. This work was supported by NBRP of China (2015CB921102)
and NSF-China under Grants Nos. 11274364 and 11574007.

\hspace{3mm}

\newpage

\begin{widetext}
\appendix
\section{Supplementary for ``Double Andreev Reflections in Type-II Weyl Semimetal-Superconductor Junctions"}

\begin{center}
Zhe Hou and Qing-Feng Sun
\end{center}

\maketitle

\section{ I. Detailed derivation of Andreev reflection coefficients $A_{1}$ and $A_{2}$ while the WSM-superconductor interface orientation angle $\theta=0$.}

While the WSM-superconductor interface orientation angle $\theta=0$, the WSM is in the $x<0$ region
and the superconductor is in the $x>0$ region.
To calculate the Andreev reflection coefficients $A_{1}$ and $A_{2}$, we first solve the BdG equation in the
type-II WSM side ($x<0$):
\begin{eqnarray}
\left(
\begin{array}{cc}
v_{1}p_{x}\sigma_{0}+v_{2} {\textbf{\emph{p}} {\cdot}\bm{\sigma}}-\mu & 0 \\
0  & -v_{1}p_{x}\sigma_{0}-v_{2} {\textbf{\emph{p}} {\cdot}\bm{\sigma}}+\mu
\end{array}
\right)
\left(
\begin{array}{cc}
f\\
g
\end{array}
\right)=
E
\left(
\begin{array}{cc}
f\\
g
\end{array}
\right),
\end{eqnarray}
where $f$ and $g$ are the electron and hole wave-functions. At a given incident energy $E$ and
the wave vectors $k_{y}, k_{z}$ in the $y,z$ direction, the eigenstates are:
\begin{eqnarray}
&&\Psi_{e+}({\bf r})=
\left(
\begin{array}{cc}
k_{+}+k_{z}\\
k_{x+}+ik_{y}\\
0\\
0
\end{array}
\right)
\exp(ik_{x+}x+ik_{y}y+ik_{z}z),
\nonumber\\
&&\Psi_{e-}({\bf r})=
\left(
\begin{array}{cc}
-k_{-}+k_{z}\\
k_{x-}+ik_{y}\\
0\\
0
\end{array}
\right)
\exp(ik_{x-}x+ik_{y}y+ik_{z}z),
\nonumber\\
&&\Psi_{h+}({\bf r})=
\left(
\begin{array}{cc}
0\\
0\\
-k_{+}'+k_{z}\\
k_{x+}'+ik_{y}
\end{array}
\right)
\exp(ik_{x+}'x+ik_{y}y+ik_{z}z),
\\
&&\Psi_{h-}({\bf r})=
\left(
\begin{array}{cc}
0\\
0\\
k_{-}'+k_{z}\\
k_{x-}'+ik_{y}
\end{array}
\right)
\exp(ik_{x-}'x+ik_{y}y+ik_{z}z),
\nonumber
\end{eqnarray}
where ${\bf r}=(x,y,z)$ and
\begin{eqnarray}
&&k_{x\pm}=\frac{v_{1}(E+\mu)\mp v_{2}\sqrt{(E+\mu)^{2}+\hbar^{2}(v_{1}^2-v_{2}^{2})(k_{y}^{2}+k_{z}^{2})}}
{\hbar(v_{1}^{2}-v_{2}^{2})},
\nonumber \\
&&k_{x\pm}'=\frac{-v_{1}(E-\mu)\pm v_{2}\sqrt{(E-\mu)^{2}+\hbar^{2}(v_{1}^2-v_{2}^{2})(k_{y}^{2}+k_{z}^{2})}}
{\hbar(v_{1}^{2}-v_{2}^{2})},
\nonumber\\
&&k_{\pm}=\sqrt{k_{x\pm}^{2}+k_{y}^{2}+k_{z}^{2}},
\nonumber \\
&&k_{\pm}'=\sqrt{k_{x\pm}'^{2}+k_{y}^{2}+k_{z}^{2}}.
\end{eqnarray}
In fact, here $k_{x-}$,  $k_{x+}$, $k'_{x+}$ and $k'_{x-}$ are the $k_x$ coordinates of the point
A, B, C and D in Fig. 5 [or in Fig.1(d) in the main text], respectively.
The wave-functions $\Psi_{e+}$ and $\Psi_{e-}$ are the eigenstates for electrons
in the tilted conduction band $E_{e+}$ and valence band $E_{e-}$.
$\Psi_{h+}$ and $\Psi_{h-}$ are the eigenstates for holes in the tilted valence
band $E_{h+}$ and conduction band $E_{h-}$ (see Fig. 5).
Both $\Psi_{e+}$ and $\Psi_{e-}$ move in the $+x$ direction, so there are two incident modes
for electrons with given energy $E$ and wave vector $k_{y}, k_{z}$.
The Andreev reflection is denoted as $A_{1}$ ($A_{2}$) if the holes are reflected in the mode $\Psi_{h+}$ ($\Psi_{h-}$).

We first consider $\Psi_{e-}$ as the incident mode. The wave functions $\Psi({\bf r})$ in the WSM
region (the $x<0$ region) can be written as follows:
\begin{eqnarray}
\Psi({\bf r}) = \Psi_{e-}({\bf r})+r_{1}\Psi_{h+}({\bf r})+r_{2}\Psi_{h-}({\bf r}),
\end{eqnarray}
where $r_1$ and $r_2$ are the reflection amplitudes to be determined by matching the boundary conditions.

\begin{figure}
\includegraphics[width=9cm]{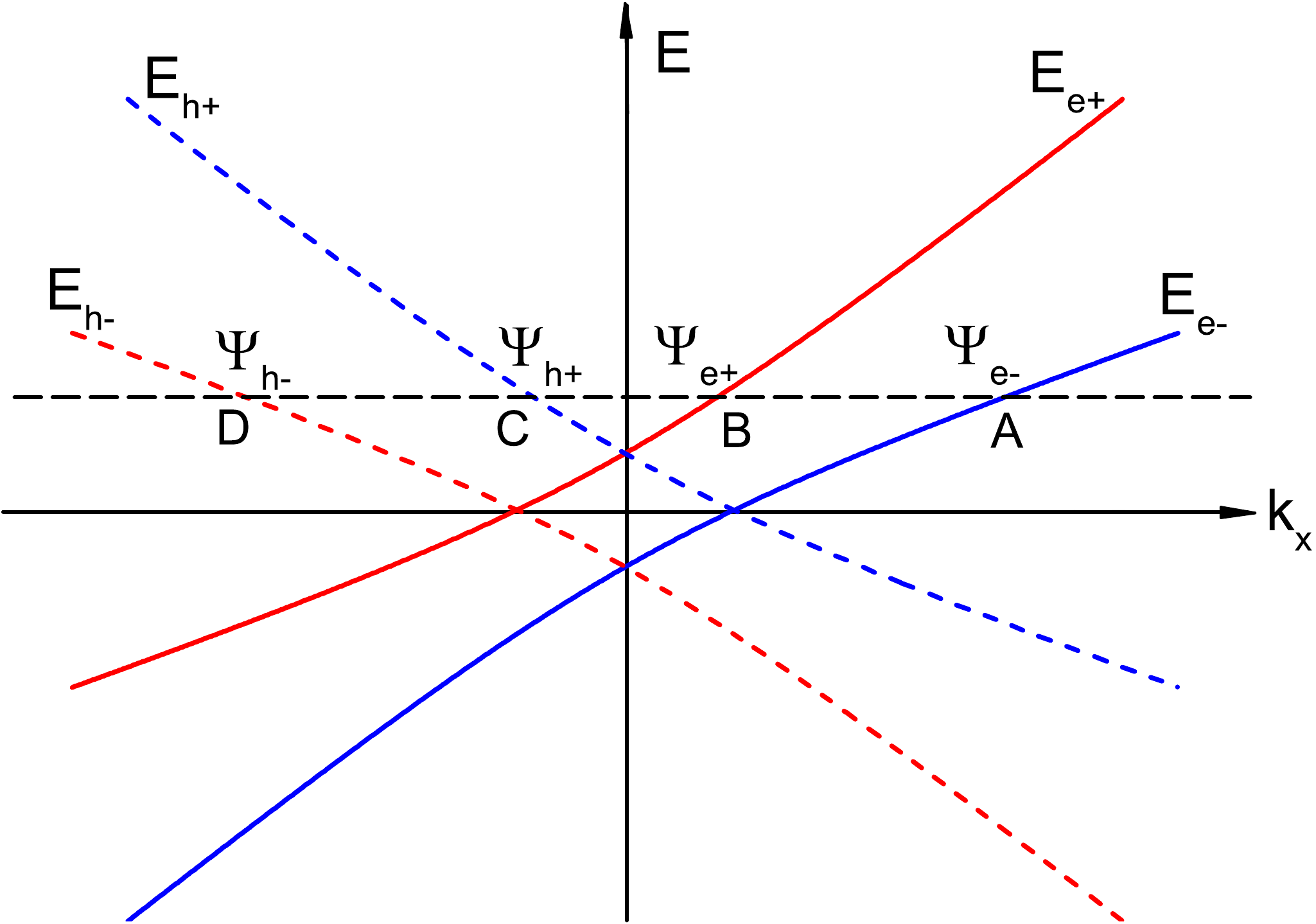}
\caption{Dispersion relations for electrons and holes in the type-II WSM. $E_{e+}$ ($E_{h-}$)
denotes the tilted conduction band for electrons (holes), and $E_{e-}$ ($E_{h+}$)
is the tilted valence band for electrons (holes).
The intersections of the black dashed line with the bands denote the incident modes $\Psi_{e\pm}$
for electrons and the reflected modes $\Psi_{h\pm}$ for holes.}
\end{figure}

The superconductor is put in the $x>0$ region with its BdG equation:
\begin{eqnarray}
\left(
\begin{array}{cc}
v_{1}p_{x}\sigma_{0}+v_{2} {\textbf{\emph{p}} {\cdot}\bm{\sigma}}-U & \Delta\sigma_{0} \\
\Delta\sigma_{0}  & -v_{1}p_{x}\sigma_{0}-v_{2} {\textbf{\emph{p}} {\cdot}\bm{\sigma}}+U
\end{array}
\right)
\left(
\begin{array}{cc}
f\\
g
\end{array}
\right)=
E
\left(
\begin{array}{cc}
f\\
g
\end{array}
\right).
\end{eqnarray}
In the large $U$ limit, the outgoing wave-functions in the superconductor are:
\begin{eqnarray}
&&\Psi_{S+}({\bf r})=
\left(
\begin{array}{cc}
e^{i\beta}\\
e^{i\beta}\\
1\\
1
\end{array}
\right)
\exp(ik_{x1}x+ik_{y}y+ik_{z}z-\tau_{1}x),
\nonumber\\
&&\Psi_{S-}({\bf r})=
\left(
\begin{array}{cc}
e^{i\beta}\\
-e^{i\beta}\\
1\\
-1
\end{array}
\right)
\exp(ik_{x2}x+ik_{y}y+ik_{z}z-\tau_{2}x),
\end{eqnarray}
where
\begin{eqnarray}
&&\beta=\left\{
\begin{array}{lll}
\arccos(E/\Delta)       &      & {\textrm{if}\quad E      <      \Delta},\\
-i\,\textrm{arcosh}(E/\Delta)       &      & {\textrm{if}\quad E     >      \Delta},
\end{array} \right.
\nonumber\\
&&k_{x1}\simeq\frac{U}{\hbar(v_{1}+v_{2})},\
k_{x2}\simeq\frac{U}{\hbar(v_{1}-v_{2})},
\nonumber\\
&&\tau_{1}=\frac{\Delta\sin{\beta}}{\hbar(v_{1}+v_{2})},\
\tau_{2}=\frac{\Delta\sin{\beta}}{\hbar(v_{1}-v_{2})},
\end{eqnarray}
The eigenstates $\Psi_{S\pm}({\bf r})$ are the superpositions of electron and hole excitations
in the superconductor, which have similar forms as the solutions
in the graphene-superconductor junction.\cite{Beenakker2}
When $E>\Delta$ these states propagate in the $+x$ direction, carrying net electron or hole current.\cite{BTK2}
When $|E|<\Delta$, $\Psi_{S\pm}({\bf r})$ both decay exponentially as $x\rightarrow +\infty$,
and the net particle current is zero, so the normal tunneling from the WSM to superconductor is prohibited.
In this case the electron incident from the WSM can only be normally reflected
as an electron or Andreev reflected as a hole.
Since no modes exist for the normal electron reflection in the Type-II WSM,
the Andreev reflection happens with unit probability.

After obtaining the modes $\Psi_{S\pm}({\bf r})$, the wave functions in the superconductor
region (the $x>0$ region) can be described as:
\begin{eqnarray}
\Psi({\bf r}) = a\Psi_{S+}({\bf r})+b\Psi_{S-}({\bf r}),
\end{eqnarray}
where $a$ and $b$ are constants to be determined by matching the boundary conditions.

The matching condition of the wave-functions at $x=0$ interface of the WSM-superconductor junction is
$\Psi({\bf r})|_{x=0^-} = \Psi({\bf r})|_{x=0^+}$. This means:
\begin{eqnarray}
 \left.\left[ \Psi_{e-}({\bf r})+r_{1}\Psi_{h+}({\bf r})+r_{2}\Psi_{h-}({\bf r})\right] \right|_{x=0^-}=
 \left.\left[ a\Psi_{S+}({\bf r})+b\Psi_{S-}({\bf r}) \right] \right|_{x=0^+}.
\end{eqnarray}
Substituting Eqs.(13) and (17) into Eq.(20), we obtain:
\begin{eqnarray}
&&r_{1}=\frac{(k_{-}-k_{z})(k_{x-}'+ik_{y})+(k_{x-}+ik_{y})(k_{-}'+k_{z})}
{(k_{+}'-k_{z})(k_{x-}'+ik_{y})+(k_{x+}'+ik_{y})(k_{-}'+k_{z})}
e^{-i\beta},
\nonumber\\
&&r_{2}=\frac{(-k_{-}+k_{z})(k_{x+}'+ik_{y})+(k_{x-}+ik_{y})(k_{+}'-k_{z})}
{(k_{+}'-k_{z})(k_{x-}'+ik_{y})+(k_{x+}'+ik_{y})(k_{-}'+k_{z})}
e^{-i\beta}.
\end{eqnarray}

\begin{figure}
\includegraphics[width=8cm]{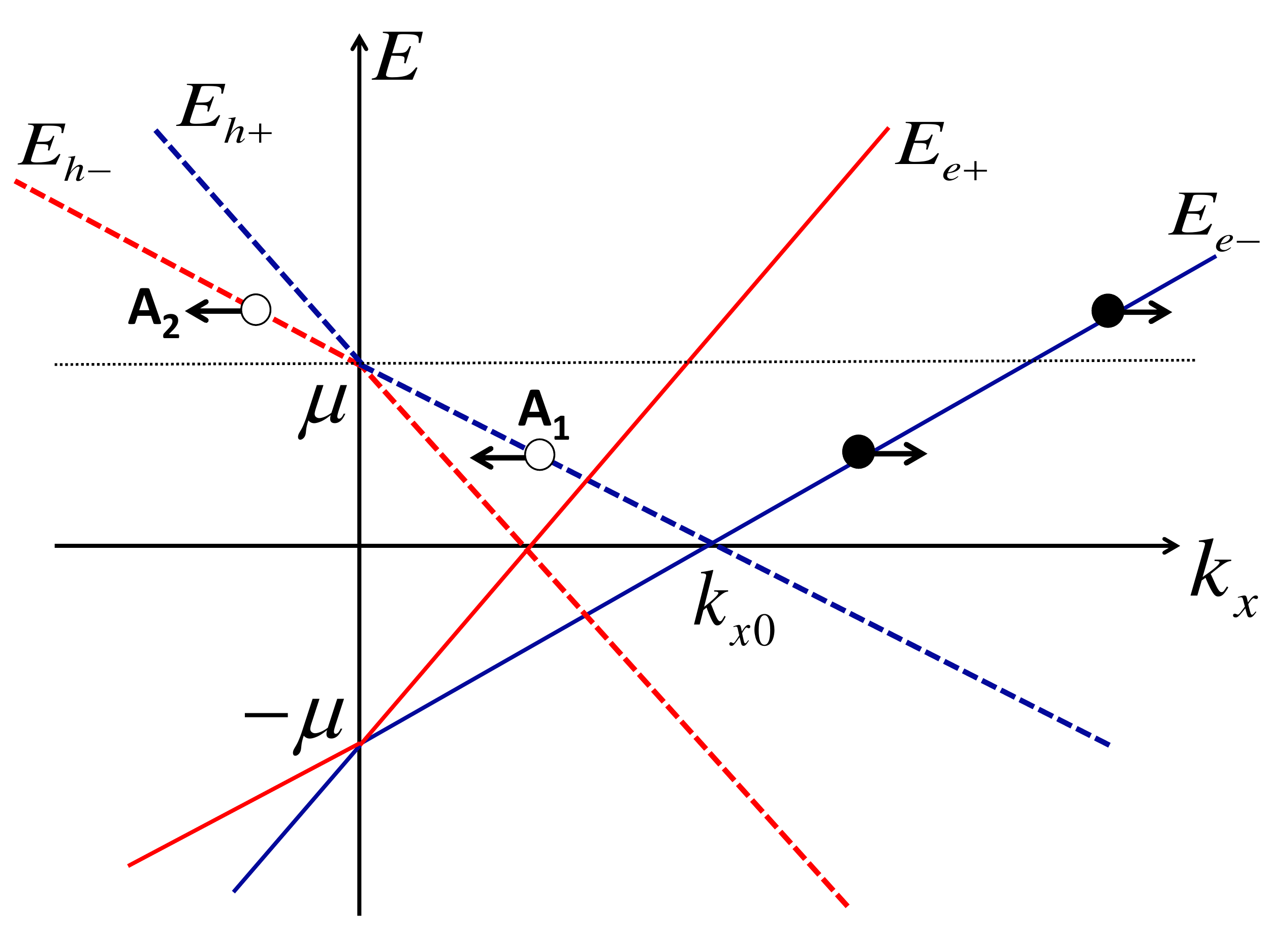}
\caption{Dispersion relation at normal incidence ($k_{y}=k_{z}=0$) with the incident angle $\alpha=0$.
The intersection of $E_{e-}$ with the $k_{x}-$axis is $k_{x0}=\frac{\mu}{\hbar(v_{1}-v_{2})}$.
The holes are reflected into the modes that are symmetric with the incident modes of electrons about $k_{x}=k_{x0}$.}
\end{figure}

The particle current density operator is defined as:
\begin{eqnarray}
\textbf{\emph{J}}\equiv \frac{1}{\hbar}[\textbf{\emph{r}},H_{BdG}]=\tau_{z}\otimes[(v_{1}\sigma_{0}+v_{2}\sigma_{x})\textbf{e}_{x}
+v_{2}\sigma_{y}\textbf{e}_{y}+v_{2}\sigma_{z}\textbf{e}_{z}],
\end{eqnarray}
where $
\tau_{z}=
\left(
\begin{array}{cc}
1 &0\\
0 &-1
\end{array}
\right)$ denotes electron-hole index, and $\textbf{e}_{x}$, $\textbf{e}_{y}$ and $\textbf{e}_{z}$
are unit vectors in $x$, $y$ and $z$ directions.
Only the $x$-component of the current density operator $J_{x}=\tau_{z}\otimes (v_{1}\sigma_{0}+v_{2}\sigma_{x})$ decides the reflection coefficients at the interface.
So the Andreev reflection coefficients $A_{1}$, $A_{2}$ can be obtained straightforwardly:
\begin{eqnarray}
A_{1}&\equiv&
\left|\frac{\langle \Psi_{h+}|J_{x}|\Psi_{h+}\rangle}{\langle \Psi_{e-}|J_{x}|\Psi_{e-}\rangle}\right||r_{1}|^{2}
=\left|\frac{(k_{+}'-k_{z})(v_{1}k_{+}'-v_{2}k_{x+}')}{(k_{-}-k_{z})(v_{1}k_{-}-v_{2}k_{x-})}\right||r_{1}|^{2}
\nonumber\\
A_{2}&\equiv& \left|\frac{\langle \Psi_{h-}|J_{x}|\Psi_{h-}\rangle}{\langle \Psi_{e-}|J_{x}|\Psi_{e-}\rangle}\right||r_{2}|^{2}=
\left|\frac{(k_{-}'+k_{z})(v_{1}k_{-}'+v_{2}k_{x-}')}{(k_{-}-k_{z})(v_{1}k_{-}-v_{2}k_{x-})}\right||r_{2}|^{2}.
\end{eqnarray}
This is Eq.(7) in the main text.
In addition, one can easily verify that when $k_{y}=k_{z}=0$ (normal incidence),
$A_1$ and $A_2$ reduce into:
\begin{eqnarray}
&&A_{1}(E)=\Theta(\mu-E)|e^{-i\beta}|^{2},
\nonumber\\
&&A_{2}(E)=\Theta(E-\mu)|e^{-i\beta}|^{2}.
\end{eqnarray}
From the band structure point of view, the modes symmetry for electrons and holes
plays an important role for Eq.(24).
Fig. 6 is the dispersion relation for the normal incidence ($k_y=k_z=0$).
One can see that the reflected holes only locate in the modes symmetric
with the incident modes of the electrons about $k_{x}=k_{x0}$.
When $E<\mu$, the reflected hole locates in the valence band of which the intra-band electron-hole
conversion results in the retro-AR ($A_{1}$).
When $E>\mu$ the symmetric reflection modes lie in the conduction band,
so $A_{1}$ disappears and $A_{2}$ arises due to the inter-band electron-hole conversion.
In addition, due to the normal incidence with the incident angle $\alpha=0$,
both the angle $\alpha_{A1}$ of the intra-band Andreev reflection and the angle $\alpha_{A2}$ of
the inter-band Andreev reflection are zero also.

\begin{figure}
\includegraphics[width=6cm]{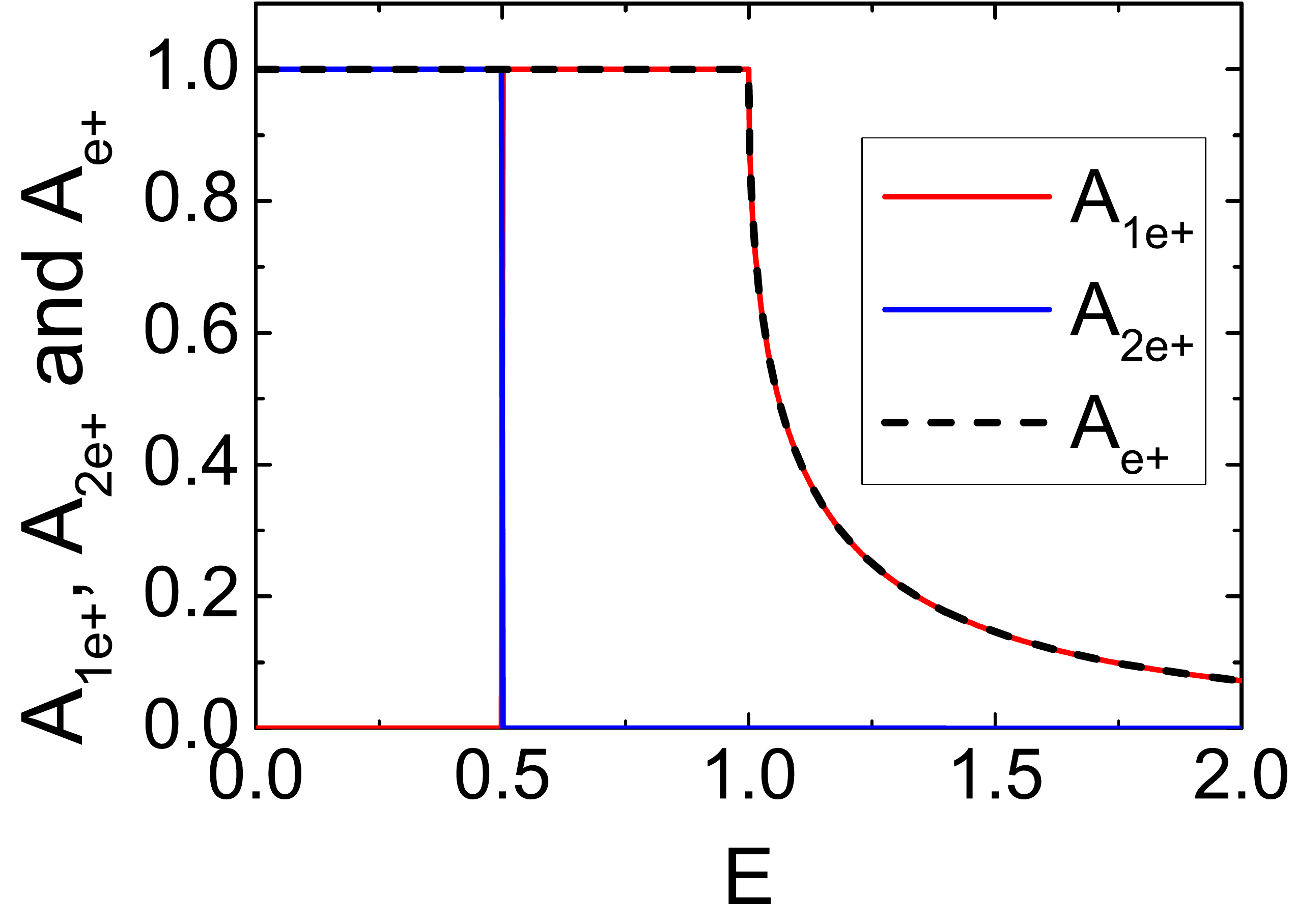}
\includegraphics[width=6cm]{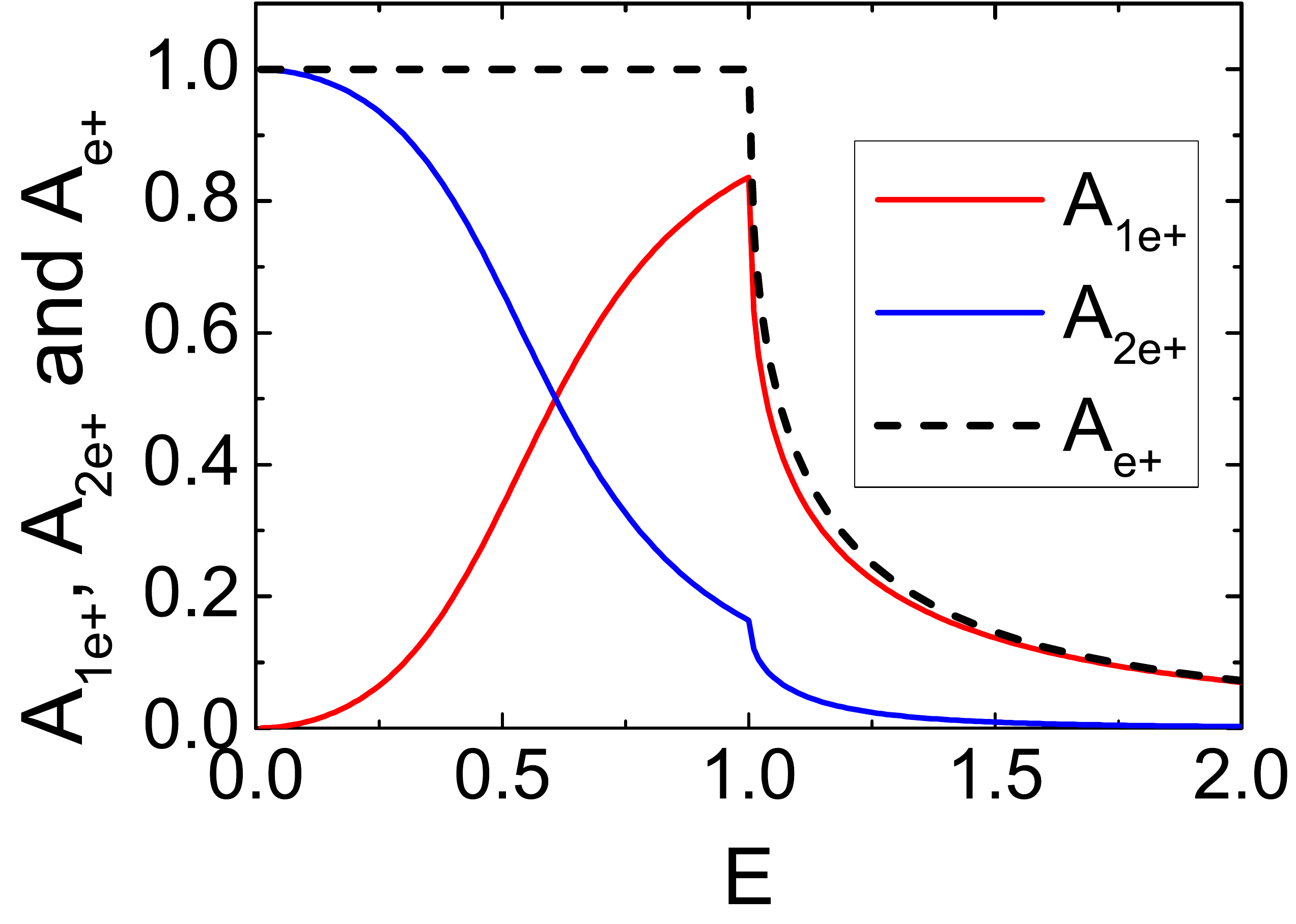}
\caption{Andreev reflection coefficients $A_{1e+}$, $A_{2e+}$ and $A_{e+}=A_{1e+}+A_{2e+}$ as functions of the incident energy $E$ for the conduction band $E_{e+}$ incidence.
The parameters are $\mu=0.5$, $v_{1}=2$, $k_{y}=0$ in (a), and $\mu=0.5$, $v_{1}=2$, $k_{y}=0.2$ in (b).
One can see that $A_{1e+}$ and $A_{2e+}$ have the same values
as $A_{2}$ and $A_{1}$ in the $E_{e-}$ incidence [see Fig.2(a) and (b) in the main text].}
\end{figure}

Up to now we have only considered the incident electron from the valence band $E_{e-}$. Similarly,
the Andreev reflection coefficients $A_{1e+}$ and $A_{2e+}$ for the incident electron from the condcution
band $E_{e+}$ can be calculated also [Here $A_{1e+}$ and $A_{2e+}$ are the coefficients for holes reflected
into the valence band $E_{h+}$ and the conduction band $E_{h-}$, respectively]:
\begin{eqnarray}
&&A_{1e+}=\left|\frac{(k_{+}'-k_{z})(v_{1}k_{+}'-v_{2}k_{x+}')}{(k_{+}+k_{z})(v_{1}k_{+}+v_{2}k_{x+})}\right||r_{1e+}|^{2}
\nonumber\\
&&A_{2e+}=\left|\frac{(k_{-}'+k_{z})(v_{1}k_{-}'+v_{2}k_{x-}')}{(k_{+}+k_{z})(v_{1}k_{+}+v_{2}k_{x+})}\right||r_{2e+}|^{2}.
\end{eqnarray}
where
\begin{eqnarray}
&&r_{1e+}=\frac{(k_{x+}+ik_{y})(k_{-}'+k_{z})-(k_{+}+k_{z})(k_{x-}'+ik_{y})}
{(k_{+}'-k_{z})(k_{x-}'+ik_{y})+(k_{x+}'+ik_{y})(k_{-}'+k_{z})}
e^{-i\beta},
\nonumber\\
&&r_{2e+}=\frac{(k_{+}+k_{z})(k_{x+}'+ik_{y})+(k_{x+}+ik_{y})(k_{+}'-k_{z})}
{(k_{+}'-k_{z})(k_{x-}'+ik_{y})+(k_{x+}'+ik_{y})(k_{-}'+k_{z})}
e^{-i\beta}.
\end{eqnarray}
The results for the conduction band $E_{e+}$ incidence are shown in Fig. 7.
One can see that $A_{1e+}$ and $A_{2e+}$ have the same values as $A_{2}$ and $A_{1}$ in
the valence band $E_{e-}$ incidence by comparison with Fig. 2(a) and (b) in the main text,
due to the unitarity of the scattering matrix.
For the conduction band $E_{e+}$ incidence, the incident direction is symmetric with that of $E_{e-}$
about the normal [see Fig. 1(d) in the main text], so $A_{1e+}$ becomes specular Andreev reflection
and $A_{2e+}$ becomes the retro-Andreev reflection.
Therefore, for given energy $E$ and wavevector $k_{y}$ and $k_z$,
no matter which incident mode the electron locates in,
the coefficient of the specular Andreev reflection (the retro-Andreev reflection) has the same values.

\vspace{5mm}

\section{ II. Detailed derivation of Andreev reflection coefficients $A_{1}$ and $A_{2}$ while the WSM-superconductor interface orientation angle $\theta\not=0$.}

In this section, we consider the case of the nonzero orientation angle ($\theta\not=0$)
of the WSM-superconductor interface.
We consider that the normal of the WSM-superconductor interface is still in the $x$-$y$ plane, but it has an intersection angle $\theta$ with the $x-$axis [see Fig. 4(a) in the main text].
If the normal of the WSM-superconductor interface is not in the $x$-$y$ plane,
one can do a rotation transformation around the $x$-axis to make the normal in the $x$-$y$ plane,
because that the Hamiltonian of the WSM in Eqs.(1) and (2) in the main text has the symmetry around the $x$-axis.
For a finite orientation angle $\theta$, the WSM is in the $x< -y \tan\theta$ region
and the superconductor is in the $x> -y \tan\theta$ region.
In this case, the incident angle $\alpha$ lies in the range $[\max(-\pi/2,\theta-\alpha_{c}),\min(\pi/2,\theta+\alpha_{c})]$ with $\alpha_{c}=\arctan(v_{2}/\sqrt{v_{1}^{2}-v_{2}^{2}})$.
For example, when $\theta=0$, the incident angle $\alpha$ lies in the range $[-\alpha_{c},\alpha_{c}]$.
$\alpha$ is restricted within $[\pi/2-\alpha_{c},\pi/2]$ for $\theta=\pi/2$.

We first take a coordinate transformation:
\begin{eqnarray}
\left(
\begin{array}{cc}
\tilde{x} \\
\tilde{y}
\end{array}
\right)
=\left(
\begin{array}{cc}
\cos\theta & \sin\theta\\
-\sin\theta & \cos\theta
\end{array}
\right)
\left(
\begin{array}{cc}
x \\
y
\end{array}
\right).
\end{eqnarray}
After the coordinate transformation, the BdG Hamiltonian becomes (for simplicity we set $p_{z}=0$):
\begin{eqnarray}
  \tilde{H}_{BdG}=
  \left(
\begin{array}{cc}
H_{+}(\tilde{\textbf{\emph{p}}})-\mu(\tilde{x}) & \Delta(\tilde{x}) \\
\Delta^{*}(\tilde{x})  &
-H_{+}(\tilde{\textbf{\emph{p}}})+\mu(\tilde{x})
\end{array}
\right),
\end{eqnarray}
where $H_{+}(\tilde{\textbf{\emph{p}}})$ is:
\begin{eqnarray}
H_{+}&=&v_{1}\sigma_{0}(\tilde{p}_{x}\cos\theta-\tilde{p}_{y}\sin\theta)
 +v_{2}\sigma_{x}(\tilde{p}_{x}\cos\theta-\tilde{p}_{y}\sin\theta)
+v_{2}\sigma_{y}(\tilde{p}_{x}\sin\theta+\tilde{p}_{y}\cos\theta),
\end{eqnarray}
where $\tilde{p}_{x}=-i\hbar\partial_{\tilde{x}}$, $\tilde{p}_{y}=-i\hbar\partial_{\tilde{y}}$.
$\mu(\tilde{x})$ and $\Delta(\tilde{x})$ in Eq.(28) are:
\begin{eqnarray}
 \mu(\tilde{x})=\left\{
\begin{array}{lll}
\mu       &      & {\tilde{x}      <      0}\\
U       &      & {\tilde{x}     >      0},
\end{array} \right. \hspace{5mm}
\Delta(\tilde{x})=\left\{
\begin{array}{lll}
0  &      & {\tilde{x}      <      0}\\
\Delta      &      & {\tilde{x}     >      0}.
\end{array} \right.
\end{eqnarray}
In the new coordinate, the WSM is in the $\tilde{x}<0$ region
and the superconductor is in the $\tilde{x}>0$ region.

Note that there are two cases need to be considered.
The first one is $v_{1}\cos{\theta}>v_{2}$, where the slopes in the $+\tilde{x}$ direction
of the tilted bands for the incident electrons (reflected holes) are all positive (negative),
and there exists double Andreev reflections at the interface.
The second case is $v_{1}\cos{\theta}<v_{2}$,
where the band tilt in the $+\tilde{x}$ direction is not so violent
that only one reflected mode exists for holes.
In this case, an Andreev reflection and a normal reflection happen at the WSM-superconductor interface,
which is similar to the normal metal-superconductor junction or graphene-superconductor junction.

Next considering an electron incidence from the WSM side to the WSM-superconductor interface
with the incident angle $\alpha$ and the incident energy $E$,
and choosing the valence band ($E_{e-}$) incidence,
the modes for incident and reflected electrons (holes) in the WSM region
($\tilde{x}<0$) can be given as follows:
\begin{eqnarray}
&&\tilde{\Psi}_{e-}=
\left(
\begin{array}{cc}
-\sqrt{\tilde{k}_{x-}^{2}+\tilde{k}_{y}^{2}}\\
(\cos{\theta}+i\sin{\theta})\tilde{k}_{x-}-(\sin{\theta}-i\cos{\theta})\tilde{k}_{y}\\
0\\
0
\end{array}
\right)
\exp(i\tilde{k}_{x-}\tilde{x}+i\tilde{k}_{y}\tilde{y}),
\nonumber\\
&&\tilde{\Psi}_{h+}=
\left(
\begin{array}{cc}
0\\
0\\
-\sqrt{\tilde{k}_{xh+}^{2}+\tilde{k}_{y}^{2}}\\
(\cos{\theta}+i\sin{\theta})\tilde{k}_{xh+}-(\sin{\theta}-i\cos{\theta})\tilde{k}_{y}
\end{array}
\right)
\exp(i\tilde{k}_{xh+}\tilde{x}+i\tilde{k}_{y}\tilde{y}),
\nonumber\\
&&\tilde{\Psi}_{h-}=
\left(
\begin{array}{cc}
0\\
0\\
\sqrt{\tilde{k}_{xh-}^{2}+\tilde{k}_{y}^{2}}\\
(\cos{\theta}+i\sin{\theta})\tilde{k}_{xh-}-(\sin{\theta}-i\cos{\theta})\tilde{k}_{y}
\end{array}
\right)
\exp(i\tilde{k}_{xh-}\tilde{x}+i\tilde{k}_{y}\tilde{y}),
\ \textrm{if} \ v_{1}\cos{\theta}>v_{2},
\nonumber\\
&&\tilde{\Psi}_{r}=
\left(
\begin{array}{cc}
-\sqrt{\tilde{k}_{xr}^{2}+\tilde{k}_{y}^{2}}\\
(\cos{\theta}+i\sin{\theta})\tilde{k}_{xr}-(\sin{\theta}-i\cos{\theta})\tilde{k}_{y}\\
0\\
0
\end{array}
\right)
\exp(i\tilde{k}_{xr}\tilde{x}+i\tilde{k}_{y}\tilde{y}),
\ \textrm{if} \ v_{1}\cos{\theta}<v_{2},
\end{eqnarray}
where the wave vectors for the incident electron and the reflected holes (electron) are
\begin{eqnarray}
&&\tilde{k}_{y}=\frac{(E+\mu)(v_{1}\sin{\theta}-v_{0}\sin{\alpha})}
{\hbar[v_{1}v_{0}\cos(\theta-\alpha)-(v_{1}^{2}-v_{2}^{2})]},
\nonumber\\
&&\tilde{k}_{x-}=\frac{(E+\tilde{\mu})\tilde{v}_{1}+
v_{2}\sqrt{(E+\tilde{\mu})^{2}+{\hbar}^{2}\tilde{k}_{y}^{2}(\tilde{v}_{1}^{2}-v_{2}^{2})}}
{\hbar(\tilde{v}_{1}^{2}-v_{2}^{2})},
\nonumber\\
&&\tilde{k}_{xr}=\frac{(E+\tilde{\mu})\tilde{v}_{1}-
v_{2}\sqrt{(E+\tilde{\mu})^{2}+{\hbar}^{2}\tilde{k}_{y}^{2}(\tilde{v}_{1}^{2}-v_{2}^{2})}}
{\hbar(\tilde{v}_{1}^{2}-v_{2}^{2})},
\nonumber\\
&&\tilde{k}_{xh\pm}=\frac{-(E-\tilde{\mu})\tilde{v}_{1}\pm
v_{2}\sqrt{(E-\tilde{\mu})^{2}+{\hbar}^{2}\tilde{k}_{y}^{2}(\tilde{v}_{1}^{2}-v_{2}^{2})}}
{\hbar(\tilde{v}_{1}^{2}-v_{2}^{2})},
\end{eqnarray}
and we have defined:
\begin{eqnarray}
&&v_{0}=v_{1}\cos(\alpha-\theta)-\sqrt{v_{1}^{2}\cos^{2}(\alpha-\theta)-(v_{1}^{2}-v_{2}^{2})},
\nonumber\\
&&\tilde{v}_{1}=v_{1}\cos{\theta},\ \tilde{\mu}=\mu+\hbar v_{1}\sin{\theta}\tilde{k}_{y}.
\end{eqnarray}
In Eq.(31), $\tilde{\Psi}_{e-}$ is the incident mode for electron.
The reflected mode $\tilde{\Psi}_{h+}$ for hole always exists regardless of the angle $\theta$,
but the other reflected mode $\tilde{\Psi}_{h-}$ for hole keeps only when $v_{1}\cos{\theta}>v_{2}$.
If $v_{1}\cos{\theta}<v_{2}$ the normal reflection mode $\tilde{\Psi}_{r}$ arises, replacing the reflected hole $\tilde{\Psi}_{h-}$.
Therefore, the wave functions $\Psi({\bf r})$ in the WSM region (the $\tilde{x}<0$ region)
can be written as follows:
\begin{eqnarray}
 \Psi({\bf r}) &=& \left\{
\begin{array}{ll}
 \tilde{\Psi}_{e-} +r_1 \tilde{\Psi}_{h+} +r_2 \tilde{\Psi}_{h-} &
 \hspace{5mm}\text{if} \hspace{2mm}v_1\cos\theta >v_2, \\
 \tilde{\Psi}_{e-} +r_1 \tilde{\Psi}_{h+} +r_{NR} \tilde{\Psi}_{r} &
 \hspace{5mm} \text{if} \hspace{ 2mm} v_1\cos\theta <v_2.
\end{array}
\right.
\end{eqnarray}
Here $r_1$, $r_2$ and $r_{NR}$ are the amplitudes of the Andreev reflection $A_1$,
the Andreev reflection $A_2$ and the normal reflection, respectively.

The outgoing modes in the superconductor region (the $\tilde{x}>0$ region) are:
\begin{eqnarray}
&&\tilde{\Psi}_{S+}=
\left(
\begin{array}{cc}
e^{i\beta}\\
(\cos{\theta}+i\sin{\theta})e^{i\beta}\\
1\\
\cos{\theta}+i\sin{\theta}
\end{array}
\right)
\exp(i\tilde{k}_{x1}\tilde{x}+i\tilde{k}_{y}\tilde{y}-\tilde{\tau}_{1}\tilde{x}),
\end{eqnarray}
and
\begin{eqnarray}
\tilde{\Psi}_{S-} &= &\left\{
\begin{array}{l}
\left(
\begin{array}{cc}
e^{i\beta}\\
-(\cos{\theta}+i\sin{\theta})e^{i\beta}\\
1\\
-(\cos{\theta}+i\sin{\theta})
\end{array}
\right)
\exp(i\tilde{k}_{x2}\tilde{x}+i\tilde{k}_{y}\tilde{y}-\tilde{\tau}_{2}\tilde{x}),
\ \textrm{if} \ v_{1}\cos{\theta}>v_{2},
\\
\left(
\begin{array}{cc}
e^{-i\beta}\\
-(\cos{\theta}+i\sin{\theta})e^{-i\beta}\\
1\\
-(\cos{\theta}+i\sin{\theta})
\end{array}
\right)
\exp(i\tilde{k}_{x2}\tilde{x}+i\tilde{k}_{y}\tilde{y}-\tilde{\tau}_{2}\tilde{x}),
\ \textrm{if} \ v_{1}\cos{\theta}<v_{2}.
\end{array}\right.
\end{eqnarray}
where
\begin{eqnarray}
&&\beta=\left\{
\begin{array}{lll}
\arccos(E/\Delta)       &      & {\textrm{if}\quad E      <      \Delta},\\
-i\,\textrm{arcosh}(E/\Delta)       &      & {\textrm{if}\quad E     >      \Delta},
\end{array} \right.
\nonumber\\
&&\tilde{k}_{x1}\simeq\frac{U}{\hbar(v_{1}\cos{\theta}+v_{2})},\
\tilde{k}_{x2}\simeq\frac{U}{\hbar(v_{1}\cos{\theta}-v_{2})},
\nonumber\\
&&\tilde{\tau}_{1}=\frac{\Delta\sin{\beta}}{\hbar(v_{1}\cos{\theta}+v_{2})},\ \tilde{\tau}_{2}=\frac{\Delta\sin{\beta}}{\hbar|v_{1}\cos{\theta}-v_{2}|},
\end{eqnarray}
After obtaining the outgoing modes, the wave functions $\Psi({\bf r})$
in the superconductor region (the $\tilde{x}>0$ region) is:
\begin{eqnarray}
\Psi({\bf r}) &=& a\tilde{\Psi}_{S+}+b\tilde{\Psi}_{S-},
\end{eqnarray}
where $a$ and $b$ are constants to be determined by matching the boundary conditions.

The matching condition at the interface $\tilde{x}=0$ of the WSM-superconductor junction is
$\Psi({\bf r})|_{\tilde{x}=0^-} = \Psi({\bf r})|_{\tilde{x}=0^+}$.
For the $v_{1}\cos{\theta}>v_{2}$ case, we have:
\begin{eqnarray}
\left.\left( \tilde{\Psi}_{e-}+r_{1}\tilde{\Psi}_{h+}+r_{2}\tilde{\Psi}_{h-}\right) \right|_{\tilde{x}=0^-}
= \left.\left( a\tilde{\Psi}_{S+}+b\tilde{\Psi}_{S-}\right) \right|_{\tilde{x}=0^+},
\end{eqnarray}
and the Andreev reflection coefficients are:
\begin{eqnarray}
A_{1}=
\left|\frac{\langle \tilde{\Psi}_{h+}|\tilde{J}_{x}|\tilde{\Psi}_{h+}\rangle}{\langle \tilde{\Psi}_{e-}|\tilde{J}_{x}|\tilde{\Psi}_{e-}\rangle}\right||r_{1}|^{2},\quad
A_{2}=
\left|\frac{\langle \tilde{\Psi}_{h-}|\tilde{J}_{x}|\tilde{\Psi}_{h-}\rangle}{\langle \tilde{\Psi}_{e-}|\tilde{J}_{x}|\tilde{\Psi}_{e-}\rangle}\right||r_{2}|^{2}.
\end{eqnarray}
On the other hand, for the $v_{1}\cos{\theta}<v_{2}$ case we have:
\begin{eqnarray}
\left.\left(\tilde{\Psi}_{e-}+r_{1}\tilde{\Psi}_{h+}+r_{NR}\tilde{\Psi}_{r}\right) \right|_{\tilde{x}=0^-}
=\left.\left( a\tilde{\Psi}_{S+}+b\tilde{\Psi}_{S-}\right) \right|_{\tilde{x}=0^+},
\end{eqnarray}
and the Andreev reflection coefficient $A_{1}$ and normal reflection coefficient $R_{N}$ are:
\begin{eqnarray}
A_{1}=
\left|\frac{\langle \tilde{\Psi}_{h+}|\tilde{J}_{x}|\tilde{\Psi}_{h+}\rangle}{\langle \tilde{\Psi}_{e-}|\tilde{J}_{x}|\tilde{\Psi}_{e-}\rangle}\right||r_{1}|^{2},\quad
R_{N}=
\left|\frac{\langle \tilde{\Psi}_{r}|\tilde{J}_{x}|\tilde{\Psi}_{r}\rangle}{\langle \tilde{\Psi}_{e-}|\tilde{J}_{x}|\tilde{\Psi}_{e-}\rangle}\right||r_{NR}|^{2},
\end{eqnarray}
where the particle current density operator is defined as $\tilde{J}_{x}\equiv \tau_{z}\otimes [v_{1}\cos{\theta}\sigma_{0}+v_{2}\cos{\theta}\sigma_{x}+v_{2}\sin{\theta}\sigma_{y}]$.
The coefficients in (40) and (42) can be numerically obtained.

\end{widetext}


\begin{thebibliography}{99}
\bibitem{Degeneracy} Here for simplicity we assume the Hamiltonian of another Weyl node with negative charity
at $\textbf{\emph{K}}'$ is $H'=v_{1}{\sigma}_{0}p_{x}-v_{2}\bm{\sigma}\cdot \textbf{\emph{p}}$, of which the ARs
coefficients $A_{1}, A_{2}$ are found to be exactly the same as those at $\textbf{\emph{K}}_{0}$.
\end{thebibliography}
\end{document}